\documentclass[sn-nature]{sn-jnl}% Math and Physical Sciences Numbered Reference Style

\usepackage{natbib}
\usepackage{multirow}%
\usepackage{amsmath,amssymb,amsfonts}%
\usepackage{amsthm}%
\usepackage{mathrsfs}%
\usepackage[title]{appendix}%
\usepackage[dvipsnames]{xcolor}%
\usepackage{textcomp}%
\usepackage{manyfoot}%
\usepackage{booktabs}%
\usepackage{algorithm}%
\usepackage{algorithmicx}%
\usepackage{algpseudocode}%
\usepackage{listings}%

\usepackage{array}
\usepackage{threeparttable}
\usepackage{booktabs}
\usepackage{array}
\usepackage{multirow}

\theoremstyle{thmstyleone}%
\theoremstyle{thmstyletwo}%

\theoremstyle{thmstylethree}%

\newcommand{\jointage}{$165^{+63}_{-34}$}

\begin{document}

\title[Article Title]{The Identification of CS\textsubscript{2} and Evidence for Carbon-Sulfur Chemical Coupling in a Warm Giant Exoplanet Atmosphere}

\author*[a]{\fnm{Anastasia Triantafillides}}\email{kearnold5@wisc.edu}
\author[a]{Thomas G.\ Beatty}
\author[b]{Matthew C.\ Nixon} % retrievals
\author[c]{Taylor J.\ Bell} % Eureka! reduction
\author[e]{Everett Schlawin} % tshirt reduction
\author[b]{Luis Welbanks} % retrievals 
\author[d]{Thomas P.\ Greene} % P.I.
\author[a]{Melinda Soares-Furtado} % Gyrochronology

%ALPHABETICAL from here onward
\author[f]{Jonathan J.\ Fortney}
\author[b]{Michael R.\ Line}
\author[g]{Nishil Mehta}
\author[b]{Sagnick Mukherjee}
\author[h]{Matthew M.\ Murphy}
\author[i]{Kazumasa Ohno}
\author[g]{Vivien Parmentier}
\author[b]{Yoav Rotman}
\author[b]{Lindsey S.\ Wiser}

\affil[a]{Department of Astronomy, University of Wisconsin–Madison, Madison, WI 53703, USA}
\affil[b]{School of Earth \& Space Exploration, Arizona State University, Tempe, AZ 85287, USA}
\affil[c]{AURA for the European Space Agency (ESA), Space Telescope Science Institute, 3700 San Martin Drive, Baltimore, MD 21218}
\affil[d]{California Institute of Technology/IPAC, 1200 E. California Blvd, MC 100-22. Pasadena, CA 91125, USA}
%\email{tgreene@ipac.caltech.edu}

\affil[e]{Astrophysics \& Space Center, Schmidt Sciences, New York, NY 10011, USA}
\affil[f]{Department of Astronomy and Astrophysics, University of California Santa Cruz, Santa Cruz, CA 95064, USA}
\affil[g]{Université Côte d’Azur, Observatoire de la Côte d’Azur, CNRS, Laboratoire Lagrange, France}
\affil[h]{Department of Physics and Astronomy, Michigan State University, East Lansing, MI 48824, USA}
\affil[i]{Division of Science, National Astronomical Observatory of Japan, Tokyo, Japan}

\abstract{Transmission spectroscopy with the James Webb Space Telescope (JWST) is revealing growing chemical complexity in giant exoplanet atmospheres. Of particular interest is sulfur, which had essentially no observational constraints before JWST. Recent work has shown that a planet's atmospheric sulfur content traces its refractory budget and is therefore a sensitive indicator of formation pathways. But despite the growing library of JWST data, the sulfur inventory of giant exoplanets remains poorly constrained: sulfur-bearing species are governed by disequilibrium chemistry and by kinetic networks that are still being revised. Here we present a transmission spectrum of the warm giant planet WASP-80 b obtained with JWST/NIRCam and MIRI over 2.4\,$\mu$m--10$\mu$m in three transits. We uniquely identify CS\textsubscript{2} in our transmission spectrum using the combination of the two absorption features in NIRCam and MIRI at a significance of $\ln (B)=17.89$ ($\sigma = 6.3$). Our grid-based retrievals yield $\mathrm{[M/H] = \:} 0.54^{+0.17}_{-0.12}$ and $\mathrm{C/O =\:}0.43^{+0.12}_{-0.08}$ which result in $\log(\mathrm{X_{CS_2}})$ abundances of $\sim-4.5$. The latest carbon-sulfur kinetics networks produce substantially greater amounts of CS\textsubscript{2} than past works, enabling good fits ($\chi^2/\mathrm{N_{data}}\sim1.2$) to the data without invoking extreme abundance patterns. These results identify CS\textsubscript{2} as an observable tracer of sulfur disequilibrium chemistry and provide observational support for theoretically predicted carbon-sulfur chemical coupling in giant exoplanet atmospheres.}

\maketitle

Transmission spectroscopy has turned giant exoplanet atmospheres into chemical laboratories in which elemental inventories, vertical mixing, and photochemistry can be inferred from trace molecular absorption features. Sulfur-bearing molecules are particularly useful because they are sensitive to all of these processes, making them diagnostic probes of atmospheric physics that are otherwise hard to observe \citep{tsai2023photochemically}. The prevalence of sulfur-bearing molecules in an atmosphere is also expected to depend strongly on its overall refractory content \citep{crossfield2025mapping}, which links to planetary formation pathways \citep{crossfield2023volatile, turrini2021tracing, pacetti2022chemical, nakazawa2026sulfur}.

Until the last several years, however, sulfur chemistry in exoplanet atmospheres remained much less explored than that of the dominant C/H/O/N species. This has changed with JWST. SO\textsubscript{2} has been identified in the transmission spectra of WASP-39 b \citep{tsai2023photochemically,powell2024sulfur}, WASP-107 b \citep{dyrek2024so2,Welbanks2024,sing2024warm}, GJ 3470 b \citep{beatty2024sulfur}, HAT-P-26 b \citep{Gressier2025hatp26b}, and HIP 67522 b \citep{thao2024featherweight} demonstrating that sulfur chemistry can be observed directly in hydrogen-rich atmospheres and that sulfur-bearing species may trace more than just thermochemical equilibrium. There is also tentative evidence for CS\textsubscript{2} and OCS -- in addition to SO\textsubscript{2} -- in the temperate sub-Neptune TOI-270 d, suggesting that carbon-sulfur species may be observable beyond the SO\textsubscript{2}-dominated cases identified so far \citep{Holmberg2024toi270,Constantinou2026}. These measurements have established sulfur as a diagnostic of atmospheric disequilibrium, but they have also exposed the problem that interpreting sulfur-bearing molecules depends strongly on the adopted kinetic network, and sulfur chemistry in exoplanet models has historically been much less mature than C/H/O/N chemistry \citep{hobbs2021sulfur,polman2023h2s,veillet2025inclusion}.

This chemical uncertainty is especially acute for CS\textsubscript{2}. In warm, methane-bearing giant planet atmospheres, earlier studies of sulfur chemistry networks generally predicted very low CS\textsubscript{2} abundances, with sulfur partitioned primarily into H\textsubscript{2}S and, under irradiation, into oxidized products such as SO\textsubscript{2} \citep{tsai2023photochemically}. This picture has been refined by recent work that showed that in cooler giant planets ($T_{eq}\lesssim600$\,K) the dominant photochemical sulfur product should switch from SO\textsubscript{2} to CS\textsubscript{2} \cite{Mukherjee2025photochem}. Further development of sulfur photochemical models using an enlarged C/H/O/N/S network predicted CH\textsubscript{4} should react to form CS\textsubscript{2}, with CH\textsubscript{2}S as the key intermediate product. Indeed, the inclusion of CH\textsubscript{2}S into photochemical networks is predicted to enable efficient carbon-sulfur coupling -- and hence greatly enhanced CS\textsubscript{2} production \citep{veillet2025inclusion, moses2024sulfur, moses2011disequilibrium}. 

The giant exoplanet WASP-80 b provides a test of this proposed CH\textsubscript{2}S carbon-sulfur coupling framework. Its atmosphere is already known to be relatively rich in CH\textsubscript{4} \citep{bell2023methane, wiser2025precise}, placing it in a chemical regime where CH\textsubscript{2}S carbon-sulfur chemistry should be especially important. This is in contrast to WASP-39 b, WASP-107 b, HIP 67522 b, and HAT-P-26 b -- the four other giant exoplanets with observed atmospheric sulfur chemistry -- which have CH\textsubscript{4} abundances more than three orders of magnitude less than WASP-80 \citep{tsai2023photochemically,Welbanks2024,sing2024warm,thao2024featherweight,Gressier2025hatp26b}. WASP-80 b also has an equilibrium temperature of $825$ K, in the middle of the temperature range where the new photochemical networks predict abundant CS\textsubscript{2}. Observationally, CS\textsubscript{2} has been tentatively reported in TOI-270 d \cite{benneke2024jwst} and while this paper was under review, was identified in the atmospheres of V1298 Tau e and TOI-6894 b \cite{dai2026photochemical, yang2026photochemical}. In both these latter cases the presence of CS\textsubscript{2} was inferred based on its prominent absorption feature near 4.6$\mu$m. Here we present JWST/NIRCam and MIRI transmission spectroscopy of WASP-80 b and uniquely identify the presence of CS\textsubscript{2} based on absorption features at 4.6$\mu$m and 6.5$\mu$m. Using a grid of self-consistent forward models, we show that the retrieved CS\textsubscript{2} abundance is consistent with the new CH\textsubscript{2}S carbon-sulfur coupling pathways, making WASP-80 b an observational validation of CH\textsubscript{2}S carbon-sulfur chemistry in giant exoplanet atmospheres.

\section{Results}

We observed two new transits of WASP-80 b on 06 October 2023 and 24 September 2023 using the NIRCam and MIRI instruments on JWST as part of the GTO programs 1185 and 1177 respectively. The NIRCam observations used the F444W filter in time-series spectroscopy mode and the MIRI observations used the LRS slitless mode. As a part of the NIRCam observations we also obtained a simultaneous broadband photometric lightcurve of the transit using the F210M filter on NIRCam's shortwave detector. JWST has previously observed the transit of WASP-80 b using NIRCam/F322W2 \citep{bell2023methane}, and we included these archival data in our analysis. In total the combined NIRCam and MIRI transmission spectrum spans wavelengths from 2.45--10.0\,$\mu$m. We binned the NIRCam data into 178 spectroscopic channels (giving a spectral resolution of $R\approx 250$) and the MIRI data into 28 spectroscopic channels (giving a spectral resolution of $R\approx 34$). 

\begin{figure*}[h!]
    \centering
    \includegraphics[scale=0.21]{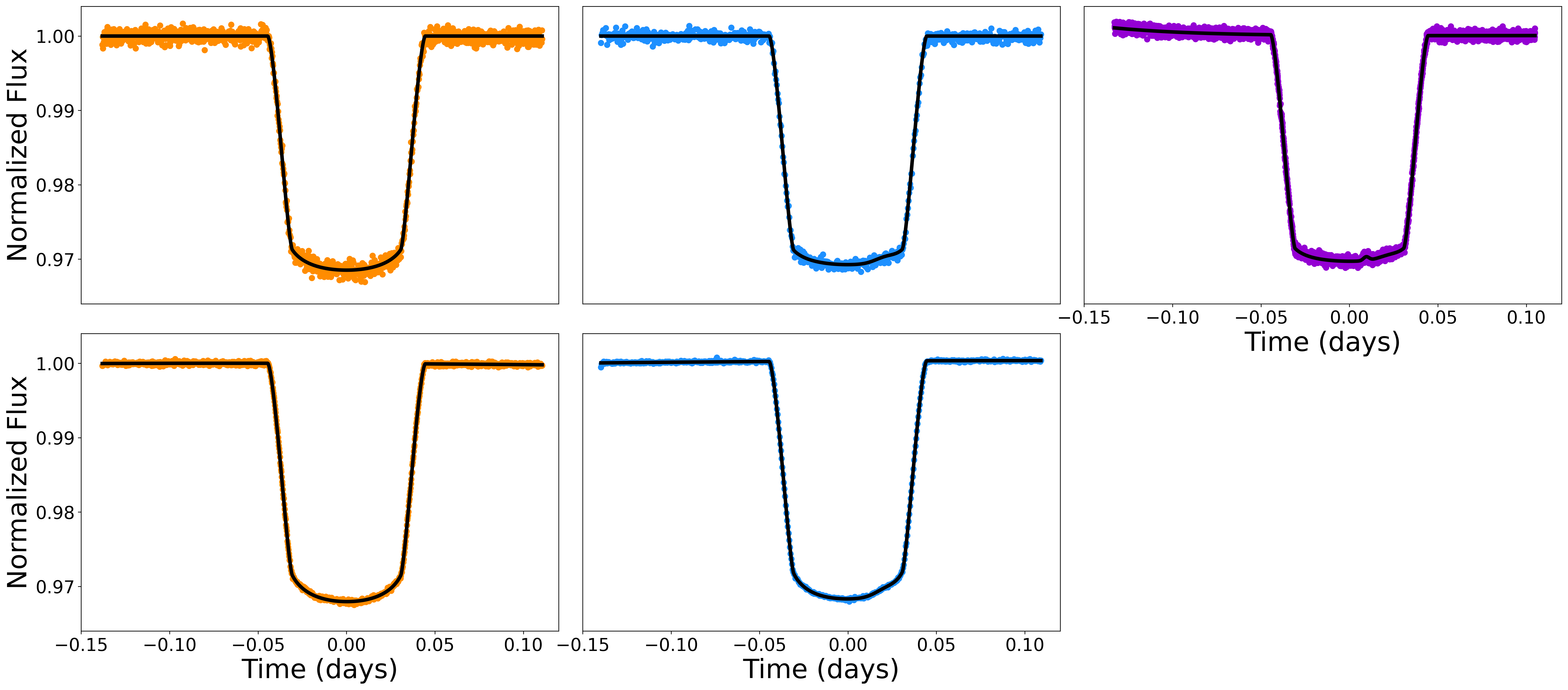}
    \caption{\textbf{Top row:} Broadband lightcurves from NIRCam/F322W2, NIRCam/F444W, and MIRI/LRS with best fit transit models plotted overtop and residuals below. \textbf{Bottom row:} Photometric observations taken simultaneously with the NIRCam/F322W2 and F444W observations. Each colored transit model corresponds to observations that were taken at the same time; i.e., orange for NIRCam/F322W2, light blue for NIRCam/F444W, and purple for MIRI/LRS.}
    \label{fig:broadbandlightcurve}
\end{figure*}

We first performed a simultaneous fit to broadband JWST transit lightcurves to refine the general orbital and planetary parameters of the WASP-80 system. Our results from this joint fit to the broadband photometry (Extended Data Table~\ref{tab:orbit}) are shown in Figure~\ref{fig:broadbandlightcurve} and were generally consistent with previous measurements, with the exception that we refined the scaled semi-major axis and inclination with respect to previously published orbital parameters \citep{Triaud2015}.

We measured the spectroscopic transit depths of WASP-80 b using the NIRCam and MIRI data. For these fits, we fixed all of the transit properties except for the planet-to-star radius ratio to the values determined from the joint broadband fit. 

\begin{figure*}[h!]
    \centering
    \includegraphics[scale=0.35]{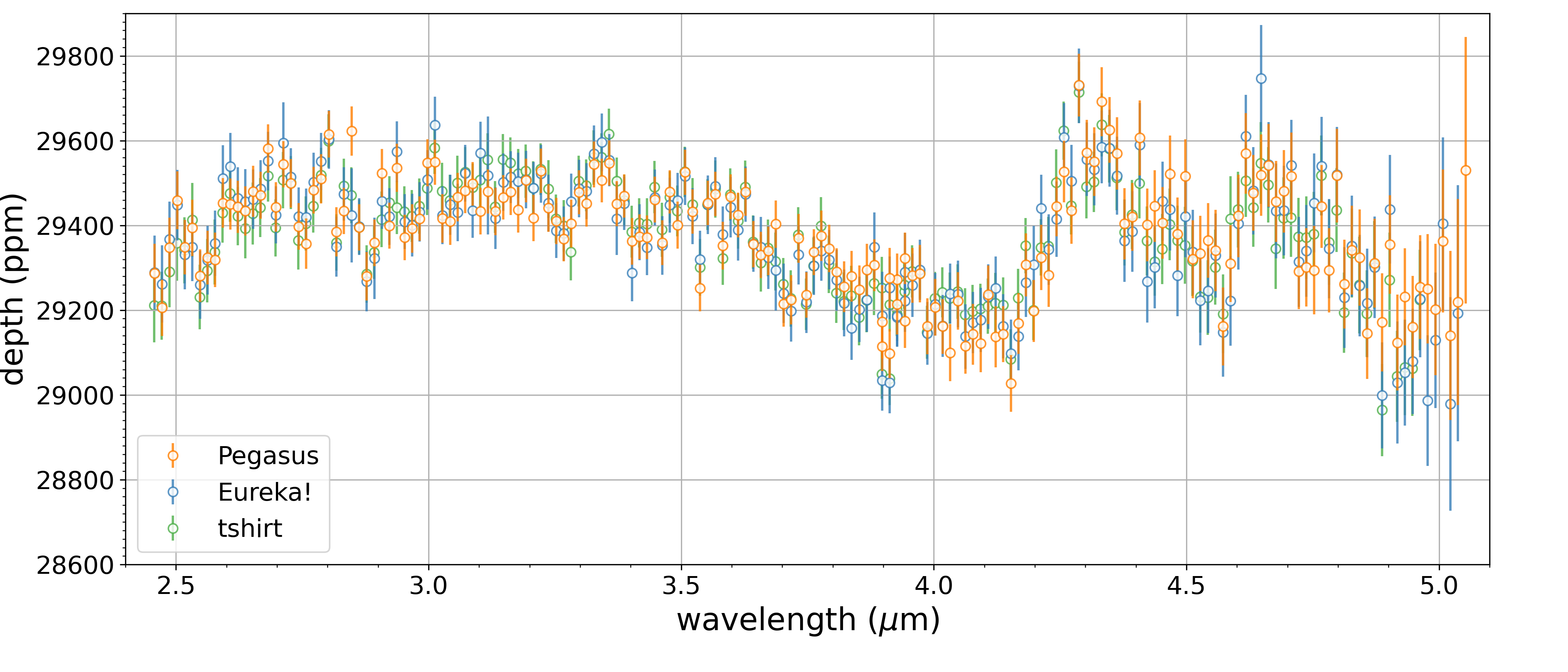}
    \caption{The three separate reductions of the NIRCam/F322W2 and NIRCam/F444W grism data. A $\chi^2$ comparison between the three reductions shows that all three are statistically consistent with each other. A more focused comparison between all three spectra near the 4.6$\mu$m CS\textsubscript{2} feature also shows no significant differences via a Bayes factor comparison.}
    \label{fig:reduction comparison}
\end{figure*}

\begin{figure*}[h!]
    \centering
    \includegraphics[scale=0.7]{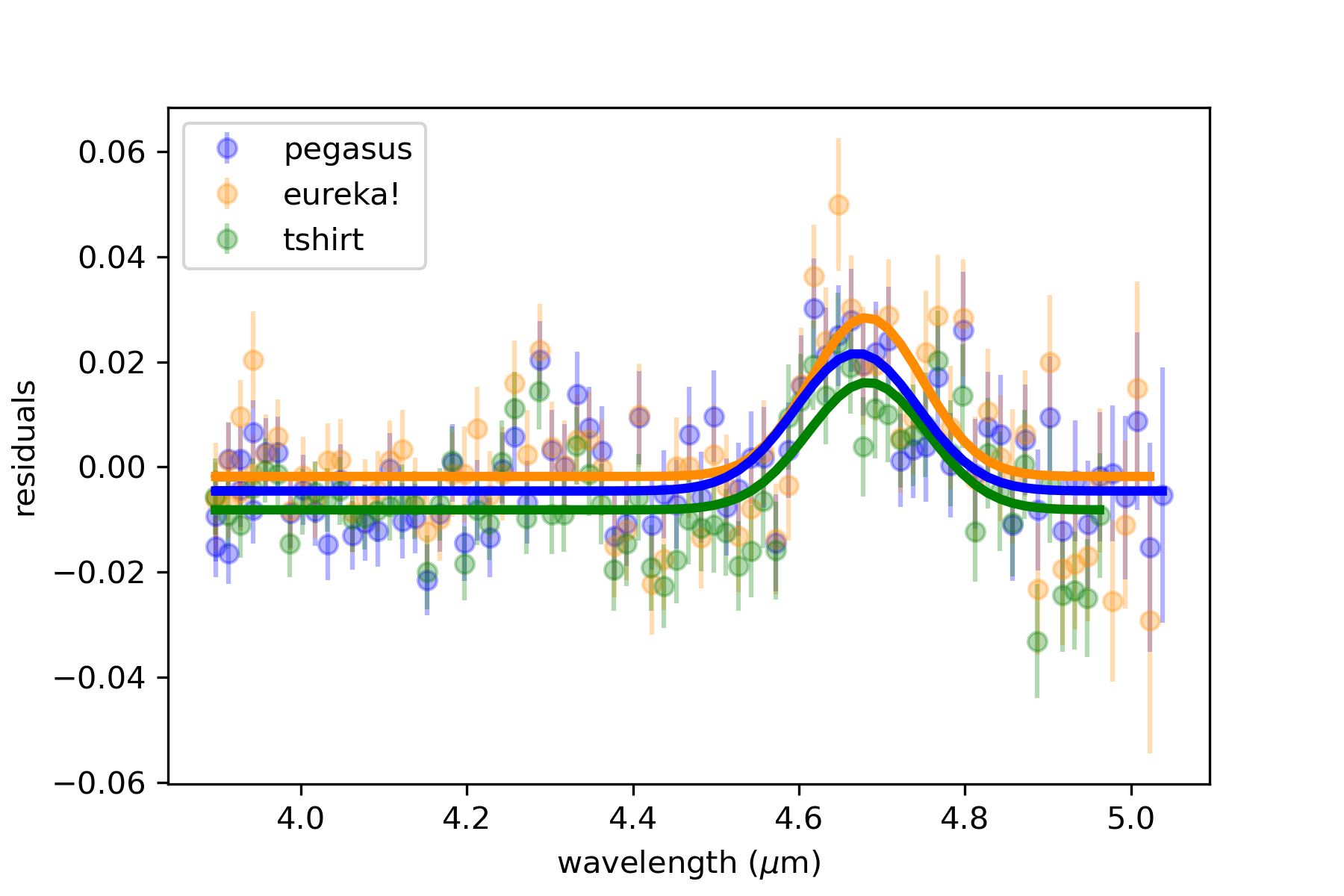}
    \caption{Residuals of all three data reductions after subtracting the best-fit forward model excluding the CS\textsubscript{2} contribution. The data in blue is Pegasus, in orange is Eureka!, and in green is tshirt. The solid lines are the best fit four-parameter Gaussian to each reductions residual feature.}
    \label{fig:gaussian comparison}
\end{figure*}

To ensure the reproducibility of our results, we performed three separate analyses of the NIRCam data using \texttt{Pegasus}\footnote{https://github.com/TGBeatty/PegasusProject}, \texttt{Eureka!} \citep{bell2022}, and \texttt{tshirt} \citep{schlawin2025tshirt} reduction pipelines (Figure~\ref{fig:reduction comparison}). Using the same system parameters, limb-darkening coefficients, and freely-fit the out-of-transit baseline, we find strong agreement between all three spectra. Figure~\ref{fig:reduction comparison}, shows a plot of all three reductions, with an offset applied to \texttt{Eureka!} and \texttt{tshirt} to match the average depth of \texttt{Pegasus}. To calculate the offset, we took a weighted average of all three reductions and applied the difference between \texttt{Pegasus} versus \texttt{Eureka!} and \texttt{Pegasus} versus \texttt{tshirt} to each respectively. We find that all three reductions give the same shape for the transmission spectra with approximately similar uncertainties. A $\chi^2$ comparison between the three reductions finds that the spectra differ by $\chi^2 /\mathrm{dof} = 0.14$ for \texttt{Pegasus} versus \texttt{Eureka!} and $\chi^2 /\mathrm{dof} = 0.12$ for \texttt{Pegasus} versus \texttt{tshirt}. In addition, we also performed a Bayesian comparison of all three spectra around the 4.6$\mu$m CS\textsubscript{2} feature using a Gaussian window check. To do so, we recalculated the best fit transmission spectrum from the retrieval analysis with the CS\textsubscript{2} abundance set to zero, subtracted this from all three spectra, and then fit a four-parameter Gaussian feature to all three reductions  in the neighborhood of the 4.6$\mu$m feature. The difference in the resulting Bayes factor between \texttt{Pegasus} and \texttt{Eureka!} is $\Delta\ln B= 5.2$, between \texttt{Pegasus} and \texttt{tshirt} is $\Delta\ln B = 4.053$ and between \texttt{Eureka!}, and \texttt{tshirt} is $\Delta\ln B= -1.144$. This demonstrates that the reductions are mildly, but not significantly, different around the CS\textsubscript{2} feature. In Figure \ref{fig:gaussian comparison} we plot all three residual features and their corresponding Gaussian fit. We chose the \texttt{Pegasus} results as our fiducial NIRCam spectrum for our atmospheric analysis.  

In addition to the new JWST data, we also performed an analysis of the TESS lightcurves of the system to identify a possible stellar rotation signal and thus better estimate the system's age using gyrochronology. The existing literature on WASP-80 disagrees on the age of the system, with estimates ranging from $\sim100^{+30}_{-20}$\,Myr \citep{Triaud2013} to $1352\pm222$\,Myr \citep{gallet2020tatoo}. None of this previous work has had access to, or analyzed, the TESS lightcurves for the system. The TESS photometry shows a strong periodicity at $3.15\pm 0.01$ days. 
Using this rotation period, we estimate a gyrochronological age of $<195$\,Myr (68\% confidence) using the \texttt{gyro-interp} software tool \citep{Bouma2023} (see Section \ref{sec:age_determination}). Because the posterior probability remains high toward the youngest ages and does not exhibit a well-defined lower bound, we report a one-sided upper limit. This places the system at the lower end of the age ranges previously reported in the literature and provides evidence that WASP-80 likely has a high UV flux further driving the C-S photochemistry on WASP-80 b.

%We interpreted the transmission spectrum using the \textit{Aurora} free retrieval framework \citep{welbanks2021aurora}, which combines an atmospheric forward model with Bayesian parameter estimation. The resulting parameters from the free retrievals are presented in Table~\ref{tab:retrieval} along with the corresponding values of Bayesian evidence (ln $Z$) in Table~\ref{tab:evidence}. We present the results from two separate retrievals; one with possible stellar heterogeneity included as a set of additional parameters and one without. The atmospheric retrieval analysis with stellar heterogeneity found results consistent with no significant spot contamination, so we adopt the ``no heterogeneity'' model as our nominal set of results.

\begin{figure*}[h!]
    \centering
    \includegraphics[scale=0.30]{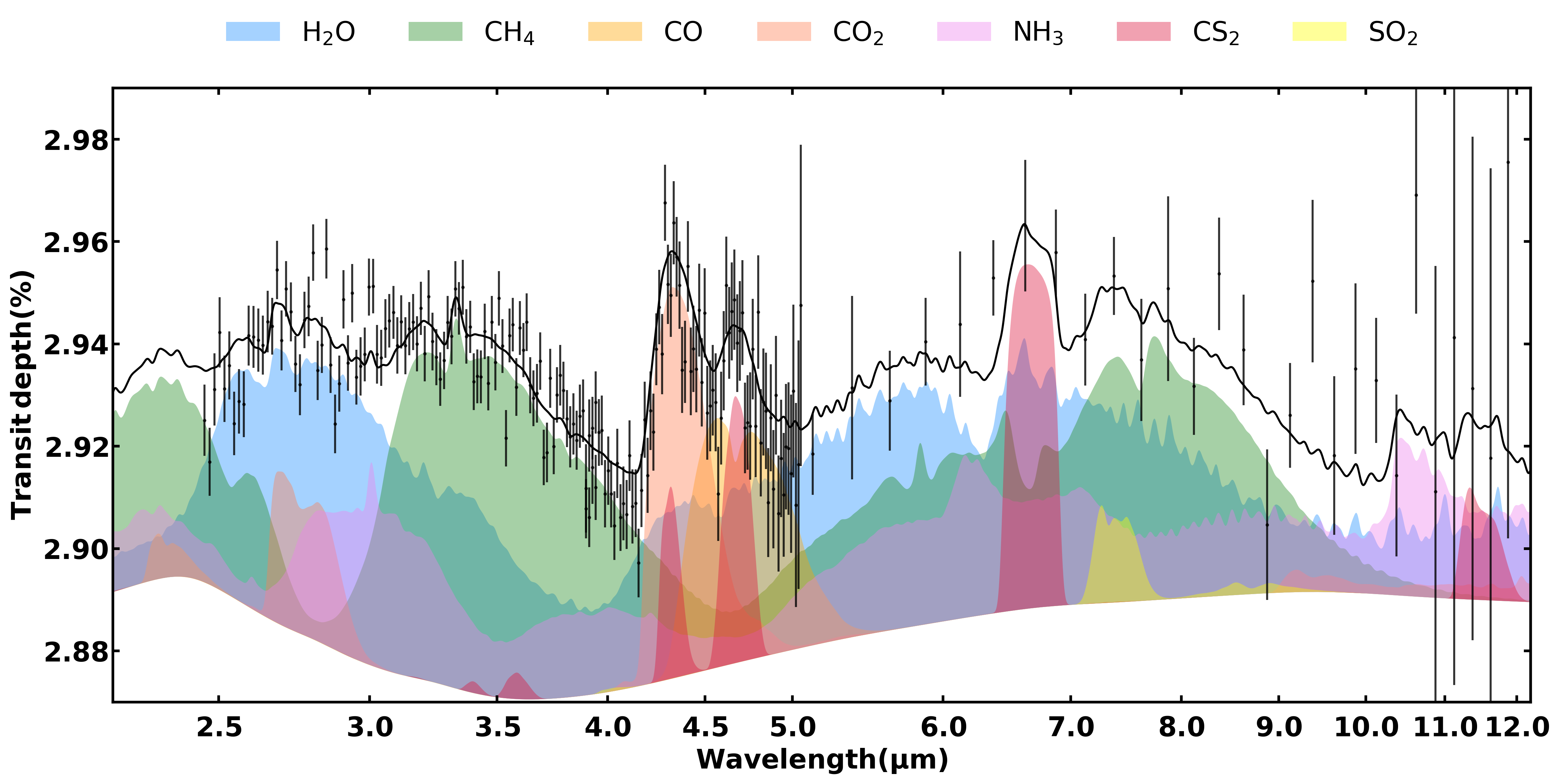}
    \caption{The best fit transmission spectrum using the results from the grid retrieval plotted as a black line, along with the transmission data from NIRCam/F322W2, F444W, and MIRI LRS plotted as black dots with 1 sigma error bars. Contributing absorption is plotted in various colors to show where gas species contribute the transmission spectrum. Based on the two absorption features at 4.6\,$\mu$m and 6.5\,$\mu$m these data uniquely identify CS\textsubscript{2} as the molecule responsible for these two features.}
    \label{fig:gas_contribution}
\end{figure*}

To interpret the observed spectrum, we use the “gridtrieval” framework \cite{bell2023methane, Welbanks2024, wiser2025precise} whereby a pre-computed grid of 1D self-consistent radiative-convective-photochemical equilibrium (1D-RCPE) model atmospheres are post-processed on the fly within a transmission spectrum nested sampling framework. To account for the newly proposed sulfur chemistry observed in WASP-80 b's transmission spectrum, we use the most recent reaction list on the VULCAN github page \citep{tsai2023photochemically} (converted to the PHOTOCHEM format), which accounts for the updated sulfur chemistry pathways \citep{veillet2025inclusion}. 
Figure \ref{fig:gas_contribution} shows the best fit to the data along with the major contributing opacity sources. The nominal best fitting solution results in a $\chi^2/\mathrm{N_{data}}$ of 1.2, suggesting that this modeling framework adequately captures the spectral shape, including the two prominent CS$_2$ features at 4.6$\mu$m and 6.5$\mu$m. The resultant log metallicity and C/O are $0.54^{+0.17}_{-0.12}$ and $0.43^{+0.12}_{-0.08}$, respectively (see Table \ref{tab:retrieval}). These constraints are consistent with those from the dayside emission spectrum \cite{wiser2025precise}. We note that a CS\textsubscript{2} abundance was not reported in the emission spectrum of WASP-80 b \citep{wiser2025precise} since that analysis did not include CS\textsubscript{2} in their atmospheric model. The general outcome of this analysis is that the self-consistent framework including the latest sulfur kinetic pathways can explain the two CS$_2$ features.

\begin{figure*}[h!]
    \centering
    \includegraphics[scale=0.23]{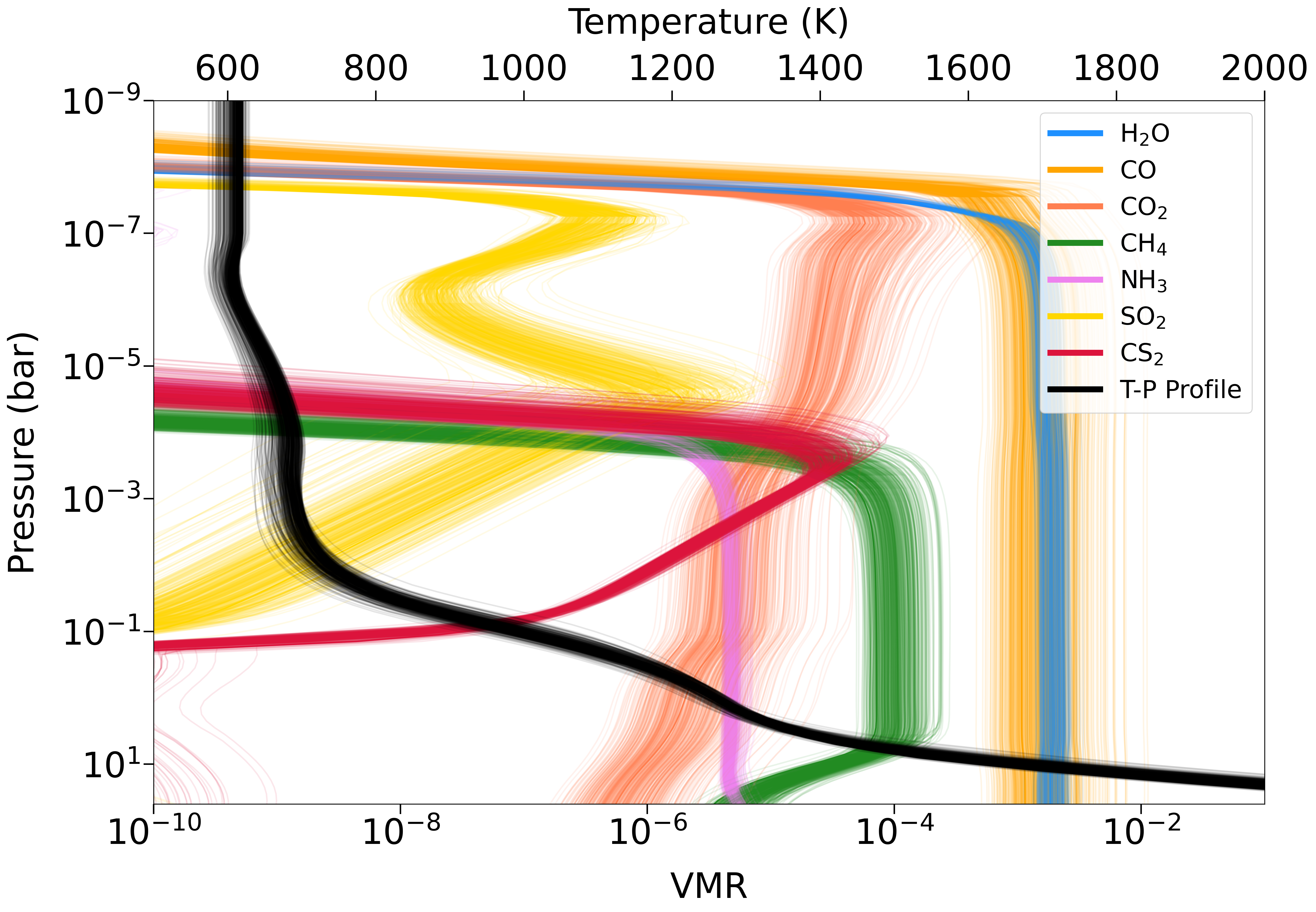}
    \caption{Random draws from the grid retrievals plotting various VMR's of major carbon-, oxygen-, hydrogen-, and sulfur-bearing species as a function of pressure. In black are the temperature-pressure profiles for each of the random draws. Note the high CS\textsubscript{2} abundance in conjunction with a very large CH\textsubscript{4} at probable pressures probed by NIRCam and MIRI transmission spectroscopy ($\sim10^{-3}$ bar.}
    \label{fig:spaghetti}
\end{figure*}

The panchromatic transmission spectrum shows prominent absorption features due to H\textsubscript{2}O\ (2.5--3.2\,$\mu$m), NH\textsubscript{3} (2.8--3.1$\mu$m), CH\textsubscript{4}\ (3.2--3.5$\mu$m and 7.5--8.0$\mu$m), CO\textsubscript{2}\ (4.2--4.5$\mu$m), and CS\textsubscript{2} (4.6--4.8\,$\mu$m and 6.3--6.8\,$\mu$m) (Figure~\ref{fig:gas_contribution}). From this, the nominal retrieval yields a CS\textsubscript{2} abundance of $\log_{10} \mathrm{X}_{\mathrm{CS_2}} \approx -4.5$ (see Figure \ref{fig:spaghetti}).

What makes our detection of CS\textsubscript{2} stand out is that it is driven by two absorption features; the one near 4.6\,$\mu$m in NIRCam and a second near 6.5\,$\mu$m in MIRI. If we include both NIRCam and MIRI in the fit, the models including CS\textsubscript{2} are significantly preferred over those without at $\ln (B)=17.89$ and $\sigma = 6.3$. The detection significance of CS\textsubscript{2} with only the NIRCam data is $\ln{(B) = 8.06}$ and $\sigma = 4.4$, and the significance using only the MIRI data is $\ln{(B)=2.31}$ and $\sigma = 2.7$. Thus while CS\textsubscript{2} is significantly detected in our NIRCam data, it is the combination of the two absorption features in NIRCam and MIRI that allow us to uniquely identify CS\textsubscript{2} as the molecule responsible. This is in comparison to the other CS\textsubscript{2} detections in the literature, all of which are less significant at $\sigma < 6$ and which need to infer the presence of CS\textsubscript{2} based on the single feature at 4.6$\mu$m.

Compared to the previously published NIRCam/F322W2 only spectrum \citep{bell2023methane}, this new data set spans a wider wavelength range allowing us to better constrain the presence of the major carbon-, oxygen, and hydrogen-bearing absorbers; H\textsubscript{2}O, CH\textsubscript{4}, CO\textsubscript{2}, and NH\textsubscript{3} in addition to the new detection of CS\textsubscript{2}.

%We also find higher abundances of H\textsubscript{2}O, CO\textsubscript{2}, CH\textsubscript{4} and NH\textsubscript{3} compared to the values measured from the eclipse spectrum of WASP-80 b \citep{wiser2025precise}, which causes our estimated metallicity for the atmosphere to be correspondingly higher at M/H = $110^{+35}_{-33}$ and the carbon-to-oxygen ratio to be C/O = $0.08^{+0.08}_{-0.04}$. 

%\begin{figure}[t!] 
%\begin{center}                                 
%\includegraphics[width=1.0\textwidth]{vmr_comparison.png}
%\end{center}
%\vskip -0.2in 
%\caption{A comparison of our retrieved abundances (points with error bars) to photochemical model predictions for WASP-80 b (solid lines). The volume mixing ratio profiles shown are from \texttt{VULCAN} simulations and use models without CH\textsubscript{2}S carbon-sulfur coupling (left) and with CH\textsubscript{2}S carbon-sulfur coupling included (right). The points showing our retrieved abundances are plotted at their representative pressure in the transmission photosphere. Most major species are reproduced at a comparable level in both models, but the predicted CS\textsubscript{2} abundance differs dramatically: the model without CH\textsubscript{2}S carbon-sulfur coupling under-predicts CS\textsubscript{2} by several orders of magnitude, whereas the CH\textsubscript{2}S coupled network yields a CS\textsubscript{2} abundance broadly consistent with the observations.}
%\label{fig:photochem}  
%\end{figure}

\section{Discussion}

The high inferred abundance of CS\textsubscript{2} on WASP-80 b is in tension with older sulfur photochemistry networks, which generally predict very little CS\textsubscript{2}, but is in agreement with recent theoretical work that incorporates carbon-sulfur coupling into photochemical models. 
%We simulated the atmosphere of WASP-80 b using the \texttt{VULCAN} photochemistry code \cite{tsai2021vulcan} using two different photochemistry networks: 
Earlier sulfur networks \citep{tsai2023photochemically} predict a negligible CS\textsubscript{2} abundance of $\log_{10}\mathrm{X_{CS_2}}\approx-9.0$, while a newer network that incorporated a CH\textsubscript{4} to CS\textsubscript{2} formation pathway by the inclusion of CH\textsubscript{2}S \citep{veillet2025inclusion} predicts a high enough VMR to be observed in transit. 

This indicates that the inclusion of the CH\textsubscript{4} to CS\textsubscript{2} formation pathway is necessary to replicate the order-of-magnitude CS\textsubscript{2} abundance we observe in WASP-80 b's atmosphere. This result mirrors more general studies of the effects of including carbon-sulfur coupling \citep{veillet2025inclusion}, which found that a fully coupled C/H/O/N/S network raises CS\textsubscript{2} abundances by one to several orders of magnitude relative to previous chemical models in methane-bearing atmospheres, while the dominant sulfur molecular reservoirs in hotter, more oxidizing cases remain H\textsubscript{2}S and SO\textsubscript{2}.

The physical origin of this difference is the coupling between reduced carbon (a carbon molecule with a low oxidation state) and sulfur chemistry. In the newly proposed coupled C/H/O/N/S networks \citep{veillet2025inclusion}, CS\textsubscript{2} can be produced through a methane-fed pathway in which carbon from CH\textsubscript{4} passes through an intermediate reaction involving CH\textsubscript{2}S before forming CS\textsubscript{2}. The net reaction to produce CS\textsubscript{2} is CH\textsubscript{4} + S$_2$ $\rightarrow$ CS\textsubscript{2} + 2H\textsubscript{2} \citep{veillet2025inclusion}. This pathway, with the inclusion of CH\textsubscript{2}S, is absent from the previous sulfur chemistry networks \citep{tsai2023photochemically}, which effectively decouples sulfur from the methane-rich carbon network and therefore under predicts CS\textsubscript{2} in a WASP-80 b-like atmosphere. Our detection provides direct observational evidence that this specific carbon-sulfur coupling must be included in models of warm giant exoplanet atmospheres. In addition, based on the net reaction described above, this also indicates that CH\textsubscript{4} is a prerequisite for substantial CS\textsubscript{2} production. Without a large reservoir of reduced carbon, the CH\textsubscript{2}S-mediated pathway cannot operate efficiently \cite{veillet2025inclusion}, and sulfur chemistry should instead remain dominated by H\textsubscript{2}S, SO\textsubscript{2}, and sulfur allotropes. CS\textsubscript{2} on WASP-80 b is therefore a tracer of the interaction between methane chemistry and sulfur photochemistry in warm, H\textsubscript{2}-rich atmospheres.

More broadly, the evidence for CS\textsubscript{2} combined with the lack of evidence for SO\textsubscript{2} provide a joint constraint suggesting that sulfur chemistry in warm giant atmospheres is shaped not only by the temperature-metallicity boundary captured by the suggested SO\textsubscript{2} ``shoreline'' \citep{crossfield2025mapping}, but also by the availability of reduced carbon. As an example, both WASP-39 b and WASP-107 b show SO\textsubscript{2} features and no CS\textsubscript{2}. This is likely because their measured CH\textsubscript{4} abundances \citep{tsai2023photochemically,Welbanks2024,sing2024warm} are at least three orders of magnitude lower than that of WASP-80 b -- indicating that even when sulfur photochemistry is active, a lack of abundant methane suppresses the CH\textsubscript{2}S carbon-sulfur coupling pathway needed to produce an observable CS\textsubscript{2} feature. The SO\textsubscript{2} ``shoreline'' may therefore be better understood as a multi-dimensional boundary. The temperature and metallicity all help determine whether sulfur chemistry is significant, but the CH\textsubscript{4} abundance can determine whether that chemistry appears primarily in oxidized species (SO\textsubscript{2}) or also includes reduced carbon-sulfur species (CS\textsubscript{2}). 

Two caveats are worth noting. First, existing CS\textsubscript{2} opacity databases are still limited for exoplanet applications. Current analyses rely primarily on room-temperature Ames/HITRAN cross-sections rather than a dedicated hot line list optimized for warm giant atmospheres \citep{Gordon2022,huang2024accurate}. Second, although the CH\textsubscript{2}S-mediated pathway \citep{veillet2025inclusion} explains the high CS\textsubscript{2} abundance we measure, the possible oxidation reactions of CS\textsubscript{2} in an H\textsubscript{2} atmosphere remain less secure, and further experimental and theoretical work on sulfur kinetics in H\textsubscript{2}-rich atmospheres is needed \citep{veillet2025inclusion}.

The identification of CS\textsubscript{2} in a methane-bearing giant atmosphere adds a new tracer to the growing set of molecules that can diagnose atmospheric metallicity, redox state, and disequilibrium chemistry. Observations of CS\textsubscript{2} through the use of two features (4.6\,$\mu$m in NIRCam and 6.5\,$\mu$m in MIRI) in the $\sim800$ K temperature range will be necessary to establish how prevalent CS\textsubscript{2} is at these temperatures, while improved opacities and sulfur networks will be needed to turn such detections into rigorous constraints on the chemical and formation histories of warm giant exoplanets.

\newpage

\section{Methods}\label{methods}

\subsection{Observations and Reduction}\label{sec:obs_fitting}

We observed two new transits of \mbox{WASP-80 b} using the F444W filter and grism of the Near Infrared Camera (NIRCam) and the slitless mode of the Low-Resolution Spectrometer (LRS) of the Mid-Infrared Instrument (MIRI) aboard the James Webb Space Telescope (JWST) \citep{kendrew2015mid, rieke2023performance}. Additionally, we perform a reanalysis of the transit observations that were taken with NIRCam's F322W2 filter and grism \citep{bell2023methane, greene2017lambda}. In addition to both longwave spectroscopic observations (F322W2 and F444W), NIRCam collected simultaneous photometric data with the shortwave detector (F210M filter).

We observed the NIRCam/F322W2 data on UT 2022 October 30 and consists of 1,227 integrations with 6 groups per integration, resulting in a total exposure time of 5.97 hours. We observed the NIRCam/F444W data on UT 2023 October 06 and consists of 639 integrations with 12 groups per integration, resulting in a total exposure time of 5.98 hours. We observed the MIRI/LRS observations on UT 2022 September 24 and consist of 4506 integrations with 29 groups per integration resulting in a total exposure time of 5.97 hours. The broadband and spectroscopic lightcurves from all three observations collectively span 2.4--10\,$\mu$m. In the following subsections, we describe how we generated the broadband and spectroscopic light curves using our three separate reductions and then briefly compare the results between the pipelines.

\subsection{NIRCam/F322W2 and F444W Reduction}

To ensure our results are robust, we used three separate analyses of the NIRCam/F322W2 and NIRCam/F444W using \texttt{Pegasus}, \texttt{Eureka!}, and \texttt{tshirt} reduction pipelines. In Figure~\ref{fig:reduction comparison} we show the results of each reduction for comparison. Our choice for the fiducial transmission spectra was the \texttt{Pegasus} reduction pipeline. We now explain each of our reduction pipelines below.

\subsubsection{NIRCam Pegasus Reduction}

As part of our analysis we re-reduced the NIRCam/F322W2 \citep{bell2023methane}, in addition to reducing the new NIRCam/F444W observations with a custom-built reduction. We used stage 1 of the JWST pipeline for basic calibration to go from the uncal to rateint images, using CRDS version 12.0.5 and \texttt{jwst\_1303.pmap} for F322W2 and CRDS version 11.17.15 and \texttt{jwst\_1210.pmap} for F444W. The NIRCam detector contains four separate amplifiers; therefore, we fit the background to each amplifier region separately. For each rateints integration, we perform a general background subtraction step by fitting a two-dimensional, second-order spline to each amplifier region on the detector and then repeated this process for every integration in time. To fit the background spline, we first mask out image rows 5 to 70 in F444W and rows 12 to 52 in F322W2 to avoid self-subtracting light from WASP-80. Additionally, we masked image rows 180 to 200 in F444W and 200 to 220 in the F322W2 to mask a nearby background star. We then perform a round of $3\,\sigma$ clipping on the unmasked portions of the image. We extrapolate the combined background spline for the whole image over the masked portions near WASP-80 and subtract it from the original image.

Next, we extract the broadband and spectroscopic light curves by performing optimal spectral extraction \citep{horne1986optimal} using the median frame in time as a spatial profile. We determined the pixel bounds for our target wavelength ranges of 2.45--3.95\,$\mu$m for F322W2 and 3.89--5.06\,$\mu$m for F444W using the equation from \texttt{tshirt} linked in the footnote\footnote{\href{https://github.com/eas342/tshirt/blob/master/tshirt/pipeline/instrument_specific/jwst_inst_funcs.py}{link to \texttt{tshirt} pixel bounds equation}}. Finally, we evenly binned wavelengths in increments of 0.015\,$\mu$m with a 250-integration wide aperture and mask 4$\sigma$ temporal outliers in each spectral bin. We account for partial pixel effects by multiplying the edge pixels by their respective fractional coverage. This results in a spectral resolution (R) of $\sim200$ for F322W2 and $\sim300$ for F444W. 

\subsubsection{NIRCam Eureka! Reduction}

Our \texttt{Eureka!}\ reduction of the NIRCam/F322W2 data began with the Stage 3 outputs from ref.\cite{bell2023methane}. For both the F322W2 and F444W data, our Stages 1 and 2 processing largely followed the defaults, with the exception of having increased the Stage 1's \texttt{jump} step rejection threshold to 6.0 (reducing the number of false-positives) and having turned off the Stage 2 \texttt{photom} and \texttt{extract\_1d} steps, which aren't needed for time-series observations. In Stage 3, both NIRCam datasets were cropped to the regions of interest, background-subtracted using a per-column linear fit to the pixels $\geq$14 ($\geq$15) away from the center of the spectral trace for F322W2 (F444W), and extracted using optimal spectral extraction \citep{horne1986optimal} using pixels within $\leq$9 ($\leq$5) pixels of the center of the spectral trace for F322W2 (F444W) and using the median integration to compute the spatial profile. For each integration, we computed a 1-dimensional spatial position and width of the stellar PSF by fitting a Gaussian to the 1D profile produced by taking the median over the wavelength axis. We then manually masked a small number of individual wavelength elements in our unbinned spectroscopic lightcurves to remove wavelength elements with anomalously high noise levels compared to adjacent pixels. Next, we spectrally binned our data into 0.015\,$\mu$m bins and then masked 4$\sigma$ temporal outliers in each spectral bin compared to a 20-integration wide boxcar filter. For the F444W data, we also trimmed the first 75 and last 75 integrations which exhibited otherwise poorly-fit temporal trends.

\subsubsection{NIRCam tshirt reduction}

We use the \texttt{tshirt} pipeline in a similar manner as previous JWST NIRCam \citep{ahrer2023ERSWasp39b,wiser2025precise,schlawin2024multiple,beatty2024sulfur} to produce count-rate images and extract spectra. We begin with the \texttt{\_uncal.fits} data products and run the JWST pipeline \texttt{calwebb\_detector1} stage version 1.13.4 on these uncalibrated data with CRDS version 11.17.15 and CRDS context \texttt{jwst\_1230.pmap} uniformly on the transmission data for both NIRCam/F322W2 and NIRCam/F444W. We manually set the jump threshold to 6~$\sigma$ for the NIRCam data. We skip the dark current correction step and apply a custom 1/f subtraction step instead of the \texttt{refpix} step in \texttt{calwebb\_detector1}. The applied custom 1/f step, \citep[ROEBA][]{schlawin2020noiseFloor1}, uses sky pixels from X=4 to X=637 for the F444W filter and X=1846 to X=2044 for the F322W2 filter. For the rest of the NIRCam steps in the Detector 1 stage, the default \texttt{calwebb\_detector1} steps are applied with \texttt{jwst\_1230.pmap}.
The Detector 1 stage includes a linearity correction after the reference pixel correction step.

For the spectral extractions, we trace the spectrum, rounding to the nearest pixel and use a full aperture width of 10 pixels (i.e. half width of 5 pixels) in NIRCam. We subtract the background column-by-column in NIRCam for all pixels more than 7 pixels away from the source within the range Y=5 to Y=64 and with a robust line fit for the range X=10 to X=70. We weight the pixels using covariance weighting \citep{schlawin2020noiseFloor1}, assuming a correlation coefficient of 0.08. The resulting extracted pixels are then binned to a wavelength grid spaced at 0.015\,$\mu$m for NIRCam, rounded to the nearest pixel.

\subsection{NIRCam F210M Shortwave photometry}

For the NIRCam F210M shortwave data we used \texttt{tshirt}'s reduction pipeline to reduce both shortwave observations that were taken simultaneously as a part of the F322W2 and F444W observations. This process is similar to the \texttt{tshirt} reduction of the F322W2 and F444W data, with a few exceptions. We use all sky pixels more than 100 pixels from the center of the source for ROEBA corrections. We centered the aperture at X=1794.27, Y=161.54 for the F444W observation 103 and center the aperture at X=1060.7,Y=165.9. In both cases, we use a source aperture radius of 79 pixels and a background annulus from 79 to 100 pixels.

\subsection{MIRI Reduction}

To ensure our results are robust, we used two separate analyses of the MIRI/LRS using the \texttt{Eureka!}, and \texttt{tshirt} reduction pipelines. Our choice for the fiducial transmission spectra was the \texttt{Eureka!} reduction pipeline. We will now explain both of our reduction pipelines below.

\subsubsection{MIRI Eureka! Reduction}

Our fiducial reduction of the MIRI/LRS slitless data used the open-source \texttt{Eureka!}\ package \citep{bell2022} with version 0.11.dev585+g8c288e84.d20240827, version 1.15.1 of the \texttt{jwst} package, and CRDS version 11.18.1 with CRDS context \texttt{jwst\_1274.pmap}. The \texttt{Eureka!}\ control and parameter files we used are available for download\footnote{Zenodo link to be created on final submission}. Our analysis closely followed the Eureka\_v1 methods of ref.\citep{bell2024} and the MIRI/LRS methods of ref.\citep{wiser2025precise}.

Our analysis began with the \texttt{\_uncal.fits} files. In \texttt{Eureka!}'s Stage 1, we made the following changes to the default parameter settings: we increased the jump rejection threshold to 7$\sigma$, ran \texttt{Eureka!}'s custom step to remove the previously documented 390\,Hz noise source \citep[e.g.,][]{Welbanks2024}, and ran the \texttt{firstframe} and \texttt{lastframe} steps to remove the first and last frames from each integration. In \texttt{Eureka!}'s Stage 2, we turned off the \texttt{photom} and \texttt{extract\_1d} steps. In Stage 3, we first masked 5$\sigma$ spatial-outliers in the background region and then subtracted the mean background level per-integration, per-row using pixels separated by 9 or more pixels from the center of the spectral trace. We then used optimal spectral extraction \citep{horne1986optimal} using the median integration to construct a spatial profile, and also measured the spatial position and width of the spectral trace for each integration to later use as covariates when fitting. In Stage 4, we masked 10 spectral elements which exhibited substantially higher noise levels compared to adjacent spectral elements. We then spectrally binned the data into 0.25~\,$\mu$m bins spanning 5--12\,$\mu$m. Finally, we masked 6$\sigma$ outliers in each channel compared to a 400-integration wide boxcar filter to remove any remaining temporal outliers.

\subsubsection{MIRI tshirt Reduction}

For the \texttt{tshirt} MIRI/LRS reduction we use the same process as the NIRCam reduction with a few exceptions. Similarly to NIRCam, we began with the \texttt{\_uncal.fits} data products and ran the JWST pipeline \texttt{calwebb\_detector1} stage version 1.13.4 on these uncalibrated data with CRDS version 11.17.15 and CRDS context \texttt{jwst\_1230.pmap}. We manually set the jump threshold to 7~$\sigma$.

For the spectral extractions, we trace the spectrum, rounding to the nearest pixel and use a full aperture width of 8 pixels. We weight the pixels using covariance weighting \citep{schlawin2020noiseFloor1}, assuming a correlation coefficient of 0.08. The resulting extracted pixels are then binned to a wavelength grid spaced at 0.25\,$\mu$m and rounded to the nearest pixel.

\subsection{Spot Treatment}

The NIRCam/F444W and simultaneous shortwave (F210M) observations show the signature of an occulted star-spot just before egress. Additionally, the MIRI/LRS broadband lightcurve also shows the signature of occulted star-spots just before egress. In the broadband joint transit model, we mask the occulted starspots in these lightcurves. However, for the NIRCam spectral lightcurves we fit a Gaussian shape to the signal of the occulted spots in lightcurve space and for MIRI we used the \texttt{fleck} transit model that includes starspot crossings. We do this to avoid biasing the orbital parameters in the broadband fit while also accurately characterizing the wavelength-dependent effect the occulted starspots may have on the transmission spectrum. 

\begin{figure*}
    \centering
    \includegraphics[scale=0.26]{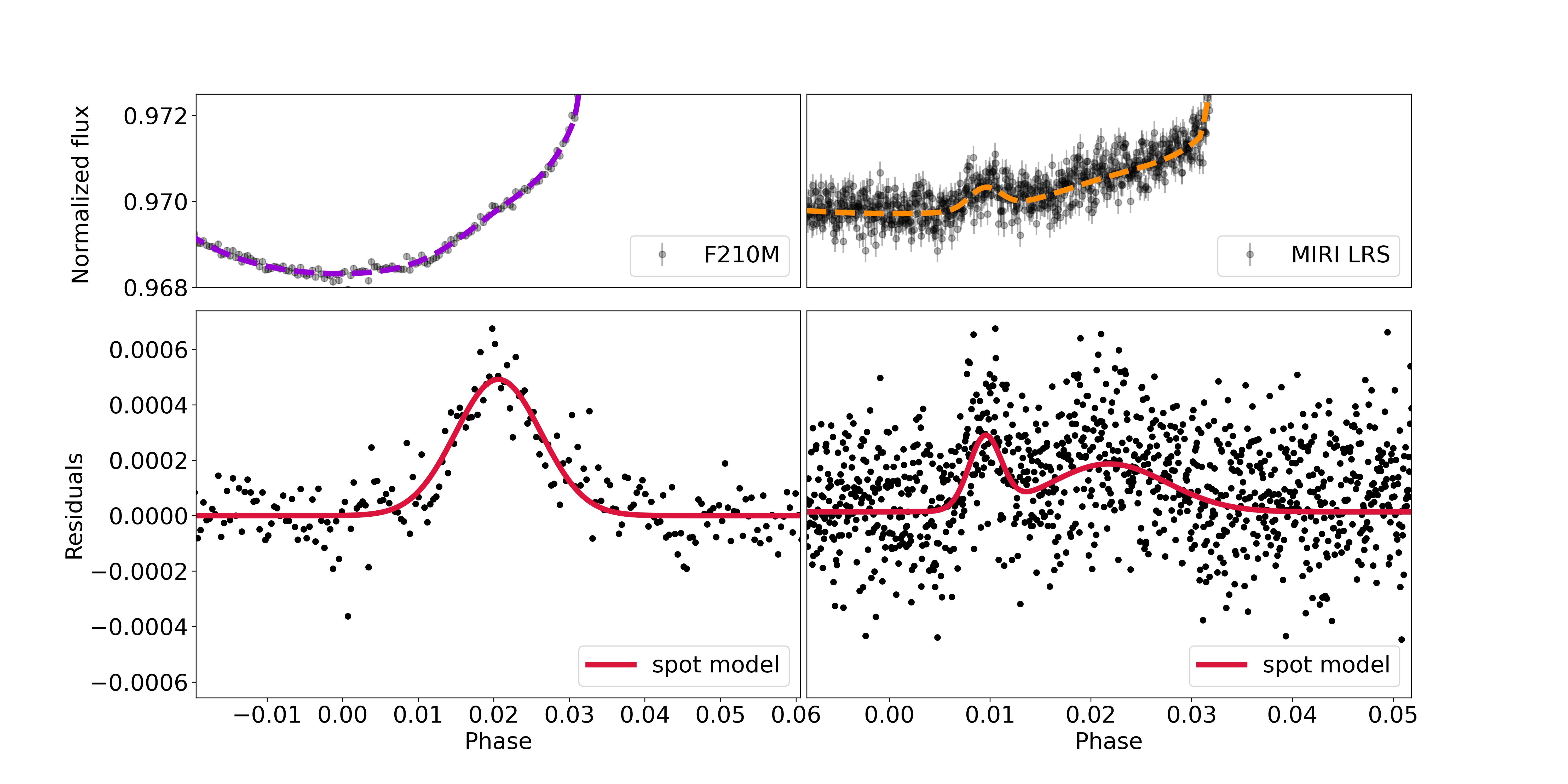}
    \caption{Top panels: Zoomed-in views of the occulted starspots in the simultaneous F210M shortwave observations (left) and in the MIRI LRS broadband observation (right). Bottom panels: Residuals between the lightcurve model and the data, with Gaussian starspot model individually fit to both lightcurves plotted in red.}
    \label{fig:spot_fit}
\end{figure*}

\begin{figure*}[h!]
    \centering
    \includegraphics[scale=0.27]{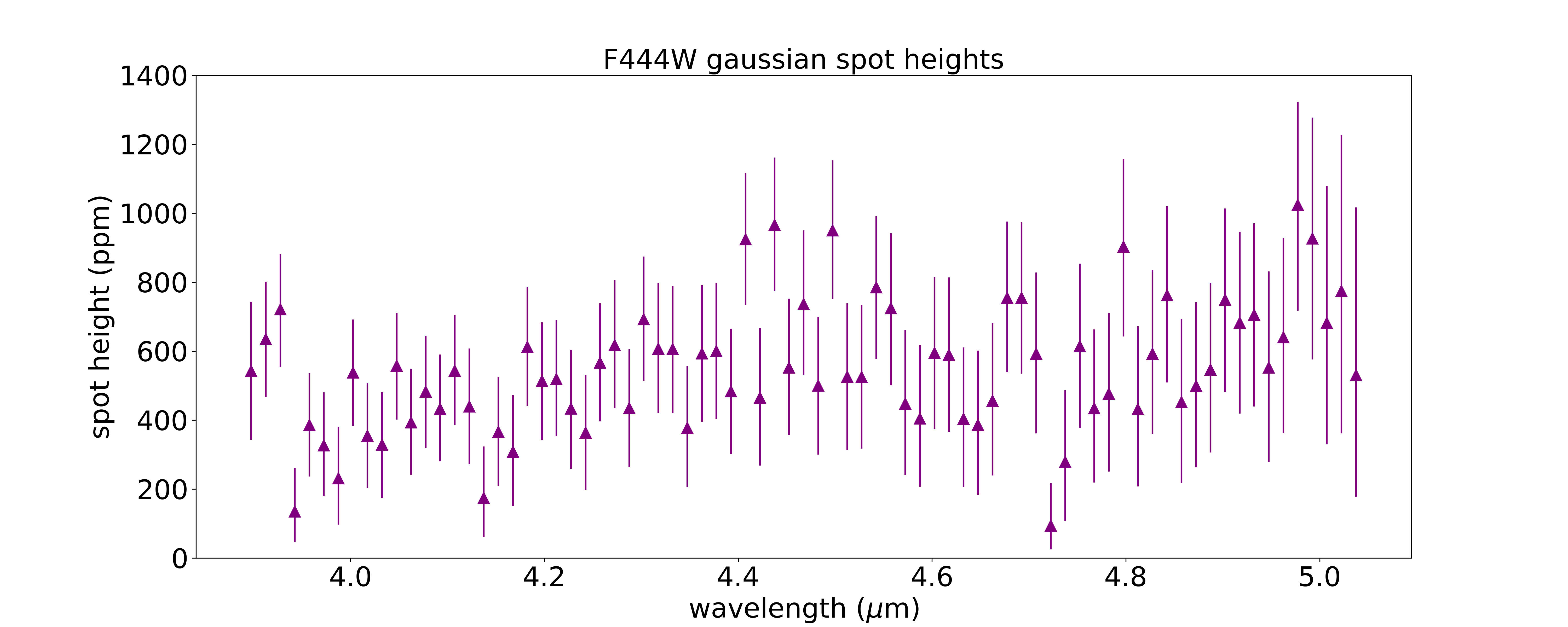}
    \caption{Retrieved heights in ppm of the Gaussian fit to the occulted starspot in the spectroscopic lightcurves.}
    \label{fig:spot heights}
\end{figure*}

We estimated the location and width of the spot-crossing events in the F444W data using its associated F210M light curve. We did so because the F210M observation has the greatest SNR for the occulted spot. For this fit we add three parameters corresponding to a Gaussian shape (spot center time, width, and height). Shown in Figure~\ref{fig:spot_fit} is the fit of the Gaussian spot to the broadband F210M observation as well as the broadband MIRI/LRS spots. The best-fit center time and width of the occulted spot from the F210M lightcurve were then fixed in the F444W spectroscopic light curve fits and we fit a single parameter for the height of the spot with uniform priors. In Figure~\ref{fig:spot heights} we plot the retrieved Gaussian spot heights from the F444W spectroscopic lightcurve fits. For our retrievals, we adopt the Gaussian spot-modeled transmission spectrum for NIRCam (as the spot's effect on the spectrum is a wavelength-independent offset) and the \texttt{fleck} spot model for MIRI. 

\subsection{Joint Broadband Fit}

Before measuring the transmission spectra of WASP-80 b, we start by simultaneously fitting a joint transit model using the broadband lightcurves from the NIRCam (F322W2, F444W, and F210M) and MIRI/LRS observations to constrain the planetary orbital parameters of the system. Assuming a circular orbit \citep{triaud2015wasp}, we fit a single set of orbital parameters for all five lightcurves and individual values for the radius ratio ($\mathrm{R_p}/\mathrm{R_*}$) for each filter. We use Gaussian priors on the period (P), time of conjunction (T$_{c}$), cosine of the inclination ($\cos i$), and the log of the scaled semi-major axis ($\log_{10}(a/R_*)$) based on the reference values from Table~\ref{tab:orbit}. Additionally, we fit uniform priors on the scaled radius of the planet ($R_{p}/R_{*}$), parameters for a linear trend in time (e.g., a slope and  y-intercept), as well as the limb-darkening coefficients ($u_1, u_2$) for each individual broadband light curve. For the MIRI/LRS broadband data, we remove the first 200 integrations to remove the worst of the initial settling behavior and include additional parameters for an exponential ramp at the beginning of the lightcurve as has been seen in other MIRI/LRS datasets \citep[e.g.,][]{bell2024}. In addition to the long-wavelength data, we include the simultaneous short-wavelength observations using the F210M filter for both the F322W2 and F444W visits. For the F210M data in the joint transit fit, we use similar Gaussian priors for the orbital parameters; however, for these F210M data we fit a quadratic trend in time, adding one extra parameter to our fit to the short-wavelength data. 

We sampled the posteriors using the \texttt{emcee} \citep{foreman2013emcee} Python package with 10 walkers, 4 chains, and a 10,000-step run with 2000-step burn-in. We confirmed that the fit had converged by ensuring the Gelman-Rubin statistic was below 1.1 for all parameters and by visually inspecting the four chains. The best-fit orbital parameters are reported in Table~\ref{tab:orbit}, and Figure~\ref{fig:broadbandlightcurve} shows the best fit model and residuals to each broadband lightcurve. 

\subsection{NIRCam Spectroscopic Fits}

Next, we performed the spectroscopic fits to the NIRCam/F322W2 and NIRCam/F444W data using three separate analysis pipelines (\texttt{Pegasus}, \texttt{Eureka!}, \texttt{tshirt}). To do so we fixed the transit center time, period, scaled semi-major axis, and the cosine of the inclination to the best fit values from the joint fit in Table~\ref{tab:orbit}. For the limb-darkening we fixed the limb-darkening coefficients to a polynomial-offset MPS2 model. Below, we describe our three separate analysis pipelines for fitting the NIRCam data.

For an initial estimate of the spectroscopic limb-darkening coefficients, we freely fit the \texttt{Eureka!}\ NIRCam/F322W2 and NIRCam/F444W spectroscopic light curves with the reparameterized quadratic coefficients \citep{kipping2013limbDarkening} and compared the fitted limb-darkening spectra to different modeled limb-darkening spectra computed by ExoTiC-LD \citep{grant2024ExoticLD}. The stellar parameters we used to compute the limb-darkening model from the MPS2 grid are T$_{\mathrm{eff}} = 4143\mathrm{\:K}$, $\log{\mathrm{(g)}} = 4.663$, and $[\mathrm{Fe/H}] = -0.13$ \citep{triaud2015wasp}. We found that applying a cubic polynomial offset (as a function of wavelength) to the $u_1$ coefficient from the MPS2 model \citep{kostogryz2022mps} provided the best fit to the overall shape and spectral features of the freely-fit limb-darkening coefficients. Meanwhile, the $u_2$ coefficient was insufficiently constrained to warrant an offset compared to the model. Such offsets in the observed limb-darkening coefficient compared to model predictions have been seen in many previous publications \citep[e.g.,][]{WASP39bERS_CO2, ahrer2023ERSWasp39b, Rustamkulov2023ERS, Feinstein2023ERS}.

\subsubsection{Pegasus NIRCam Spectroscopic Fits}

To generate the transmission spectrum, we fix the orbital parameters to the best-fit values from our joint broadband fit. We then fit the scaled planetary radius with a linear trend in time. We use the Python package \texttt{BATMAN} to model the lightcurves. Similar to the broadband fit, we sampled the posteriors using the \texttt{emcee} Python package with 10 walkers and a 10,000-step production run for each spectral channel with 2,000 step burn-in. During our fiducial spectroscopic fits we fixed the limb-darkening coefficients to the polynomial-offset MPS2 model described above.

Once we finished the spectroscopic fits, we then applied an offset correction between NIRCam/F322W2 and NIRCam/F444W. We do this by calculating the weighted average of the three points that overlap in wavelength and subtract the difference between them. Prior to running the atmospheric retrievals, this offset was applied to the NIRCam/F322W2 data. The offsets between modes was also a free fitting parameter in the model (see Table \ref{tab:retrieval} for parameter estimates).

For our data reduction and extraction, we used two additional analysis pipelines to fit the F322W2 and F444W transmission spectra as a check for self-consistency. We used the data reduction and light curve fitting from both \texttt{Eureka!} and \texttt{tshirt} to check our fiducial spectroscopic fitting, and we now describe those independent reductions.

\subsubsection{Eureka! NIRCam Spectroscopic Fits}

For NIRCam/F322W2, our astrophysical model was a \texttt{BATMAN} transit model \citep{batman} with the orbital parameters fixed to the best-fit values from our joint broadband fit. However, for NIRCam/F444W there was clear evidence for starspot crossing events during the transit, so we used the \texttt{fleck} transit model \citep{fleck2020} to model occulted starspots. Because starspot locations and sizes are largely wavelength-independent, we jointly fit the NIRCam/F444W broadband data with the simultaneous F210M shortwave photometry to get better constraints on the starspot locations.

We found that the F444W and simultaneous F210M data were best-fit by a 3-spot model, with the three occulted spots lying within or near the transit chord (which is offset from the sub-observer point on the star by $14.24_{-0.46}^{+0.30}$ degrees and has a half-width of $10.201\pm0.023$ degrees). For each spot, the inferred contrast, latitude, and radius of the spot were substantially correlated with each other at the precision of our data. To try to constrain the problem, we imposed the following priors: spot contrast, $F_{\rm spot}/F_*=\mathcal{N}(0.85, 0.2^2)$ based on the expectation for $\Delta T_{\rm spot}\sim300$\,K for $T_{\rm eff}\approx4100$\,K; spot radius ratio, $R_{\rm spot}/R_*=\mathcal{U}(0,0.3)$; latitudinal offset compared to the center of the transit chord, $\Delta\phi=\mathcal{N}(0,10^2)$ degrees; and spot longitudes of $\mathcal{N}(+18,10^2)$ degrees, $\mathcal{N}(+38,10^2)$ degrees, and $\mathcal{N}(+57,10^2)$ degrees based on initial fits that indicated these approximate spot longitudes and where 0$^{\circ}$ is the sub-observer stellar longitude and positive offsets are eastward and result in spot crossings that occur after mid-transit.

Our three fitted spots were located at longitudes of $18.3_{-1.1}^{+1.1}$ degrees east, $38.46_{-0.87}^{+0.85}$ degrees east, and $60.1_{-3.8}^{+3.6}$ degrees east of the sub-observer point. Our fitted spot contrast for F210M was $0.9696_{-0.024}^{+0.0080}$ and for the broadband F444W data was $0.9798_{-0.016}^{+0.0058}$; note that \texttt{Eureka!}'s \texttt{fleck} implementation assumes that all starspots have the same spot contrast for any given wavelength. We found that these spot contrasts could be fit with a $\Delta T_{\rm spot}\sim60$\,K \texttt{PHOENIX} model (linearly interpolating between grid points) or similar temperature blackbody model. For our NIRCam/F444W spectroscopic fits, we chose to impose a Normal prior on the spot contrast with standard deviation $\pm0.01$ and a mean centered on the predictions from our $\Delta T_{\rm spot}\simeq60$ K \texttt{PHOENIX} model, and we fixed the spot radii and locations to the maximum log-probability values from the broadband fit. For our spectroscopic fits, we fixed the limb-darkening coefficients to the polynomial-offset MPS2 model described above.

Our systematic model for both NIRCam/F322W2 and F444W was comprised of a linear trend in time and a linear decorrelation against the temporal changes in the spatial position and width of the stellar PSF computed in Stage 3. We also included a white-noise error inflation term to capture any additional white noise beyond the photon limit. We used the \texttt{dynesty} dynamic nested sampling algorithm to sample our model parameters, using an initial lives points of $\max(121, N_{\rm dim}(N_{\rm dim}+1)//2)$ (where $N_{\rm dim}$ is 6 for F322W2 and 7 for F444W, and $//$ indicates integer division), `multi' bounds, the `rwalk' sampling algorithm, and an initial convergence criterion of $d\log\mathcal{Z}<0.01$ where $\mathcal{Z}$ is the computed Bayesian evidence; batches of $\max(25, N_{\rm live,init}/2)$ additional live points were then added with pfrac=0.9 (favoring a precise posterior distribution over a precise Bayesian evidence) until the automated stopping criterion was reached. We then adopted the median of the samples as our best-fit value and used the 16th and 84th percentiles of the samples to estimate our $\pm$1$\sigma$ uncertainty intervals.

\subsubsection{tshirt NIRCam Spectroscopic Fits}

We fit all spectroscopic bins individually in two iterations of the treatment of limb darkening. In the first iteration, we fit the uninformative-prior limb darkening coefficients \citep{kipping2013limbDarkening}. In the second iteration, we scaled an ExoTiC-LD limb darkening model \citep{grant2024ExoticLD} to best fit the first iteration and then held the limb darkening coefficients fixed for our second fit.  We fit all lightcurves using a constant ephemeris (P=3.06785192079346,t$_0$=2456487.42500202,$i$=88.985, $a/R_*$ = 12.65 and $e$=0). We assumed a baseline consisting of a constant plus a linear term with time and a linear coefficient with the Focal Plane Array Housing (FPAH) temperature \citep{schlawin2024multiple}, and we masked points that were 5$\sigma$ from the model to remove cosmic rays. We also bin the time-axis to 300 points for faster evaluation.

\subsection{MIRI Spectroscopic Fits}

For the MIRI/LRS data we performed the spectroscopic fits using the \texttt{Eureka!} and \texttt{tshirt} analysis pipelines. Similar to the NIRCam fits, we fixed the transit center time, period, scaled semi-major axis, and cosine of the inclination to the best-fit values from the joint fit in Table~\ref{tab:orbit}. Additionally, we fixed the limb-darkening parameters and used the same poly-shifted MPS2 model that we had used for the NIRCam data. Below we describe the \texttt{Eureka!} and \texttt{tshirt} analysis for fitting the MIRI/LRS data.

\subsubsection{Eureka! MIRI Spectroscopic Fits}

Similar to the NIRCam/F444W data, our initial MIRI/LRS fits indicated that there were multiple starspot crossings; as a result, we used a \texttt{fleck} transit model that included starspot crossings. The systematics model for the LRS data was comprised of a linear trend in time, an exponentially decaying ramp in time, a linear decorrelation against the changes in the spatial position and width of the spectral trace, and finally a Gaussian process using a \mbox{Mat\'ern-3/2} kernel as a function of time computed using the \texttt{celerite2} package \citep{celerite2}. We also included a white-noise error inflation term to capture any additional white noise beyond the photon limit. We then used the same \texttt{dynesty} dynamic nested sampling setup as for our \texttt{Eureka!} NIRCam fitting (where $N_{\rm dim}$ is 11).

Our MIRI fits including starspots followed the same general methodology as those for the F444W+F210M fits. We again found that three occulted starspots were favored. Our fitted spot contrast was $0.970_{-0.014}^{+0.011}$ ($\Delta T_{\rm spot}\sim100$\,K) and the three spots were located at longitudes of $26.0_{-2.9}^{+3.5}$ degrees west, $23.0_{-1.4}^{+1.4}$ degrees east, and $42.1_{-1.4}^{+1.4}$ degrees east of the sub-observer point. As with our F444W spectroscopic fits, our LRS spectroscopic fits imposed a Normal prior on the spot contrast with a standard deviation of $\pm0.01$ and a mean centered on the predictions from our $\Delta T_{\rm spot}=100$\,K blackbody model (since \texttt{PHOENIX} only extends to $\sim$5.5\,$\mu$m), and we fixed the spot radii and locations to the maximum log-probability values from the broadband fit.

\subsubsection{tshirt MIRI Spectroscopic Fits}

For our fits of MIRI/LRS with \texttt{tshirt} we performed the spectroscopic fits using the same limb darkening treatment we used for the NIRCam fits with \texttt{tshirt}. Again, we fit all lightcurves using a constant ephemeris and assume a baseline that is a constant plus a linear coefficient with time and a linear coefficient with the Focal Plane Array Housing (FPAH) temperature \citep{schlawin2024multiple}. We also bin the time-axis to 300 points for faster evaluation. For MIRI, we also include an exponential settling component of the baseline with a log normal prior on the settling time centered on 70 minutes and a geometric width of 1.

\section{Determining the age of the host star WASP-80}\label{sec:age_determination}
\subsection{Measuring the Stellar Rotation Period From TESS}

To determine a more accurate age for WASP-80, we searched for signs of stellar rotation in the TESS light curves of WASP-80. TESS observed WASP-80 in Sectors 54 and 81. To conduct our search we used the measured \textsc{PDC\_SAPFLUX} in the Sector 54 TESS data produced from the SPOC pipeline. TESS observed this sector from UT 2022 July 9 to UT 2022 August 9. For simplicity, we removed from this light curve all the points that were observed during a transit of WASP-80 b. 

\begin{figure*}[h!]
    \centering
    \includegraphics[scale=0.35]{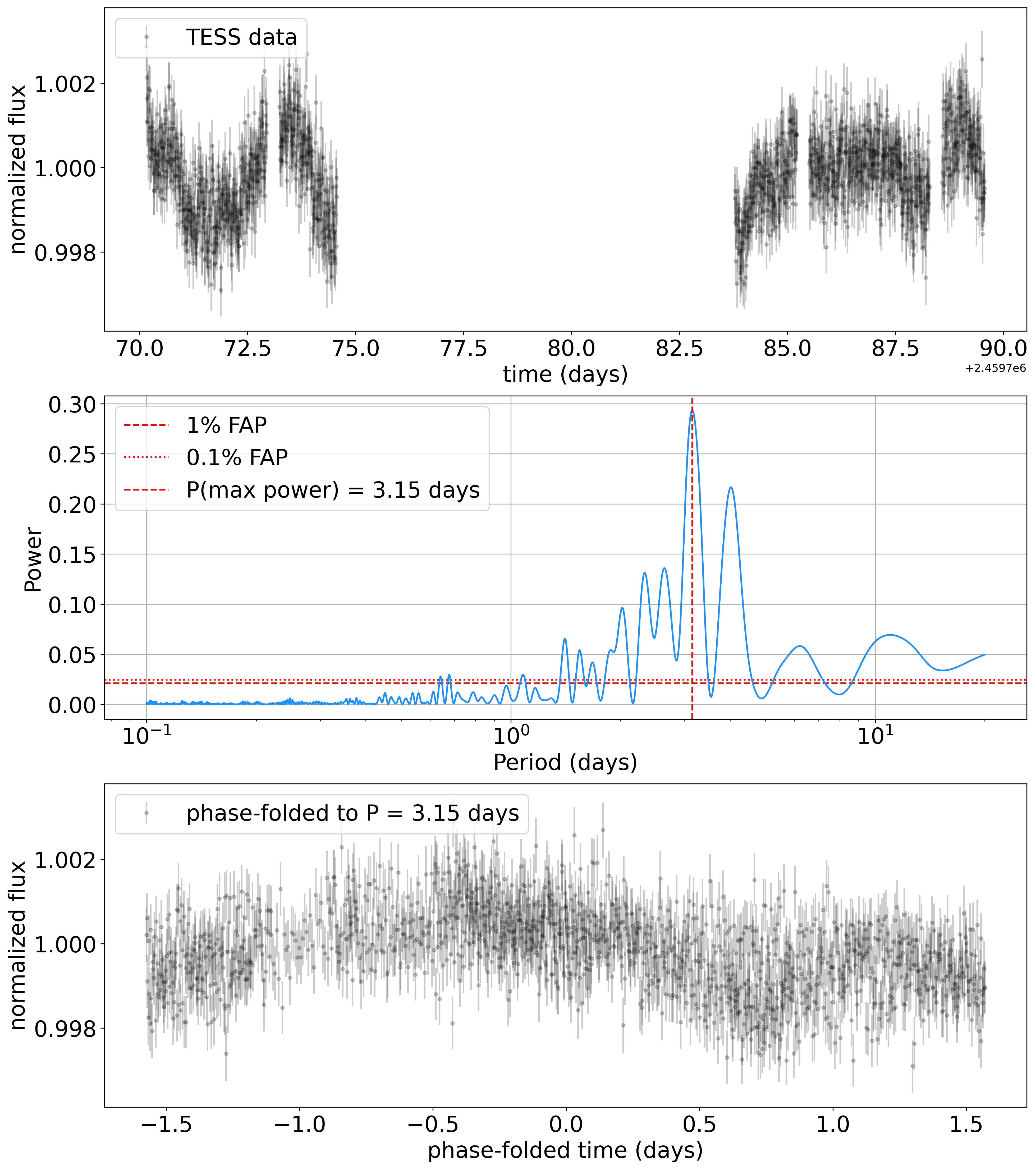}
    \caption{Top panel: phase-folded masked transit of \mbox{WASP-80 b} from TESS observations taken between UT 2022 July 9 to UT 2022 August 9. Middle panel: The Lomb-Scargle periodogram of the TESS data with the masked transit. Bottom panel: The masked TESS data phase-folded to the best period found by the Lomb-Scargle periodogram.}
    \label{fig:periodogram}
\end{figure*}

We then used a Lomb-Scargle (L-S) periodogam \citep{lomb1976least,scargle1982studies} to search for possible stellar rotation signals. We set the maximum allowed period to 20 days and the minimum period to 0.1 day. Figure~\ref{fig:periodogram} shows the results of this analysis. The L-S search identified a strong peak at $P = 3.15$ days, with a power noise threshold of $0.02$ corresponding to a $0.01\%$ false alarm probability (middle panel of Figure~\ref{fig:periodogram}). Using this period, the phase-folded light curve shows a roughly sinusoidal signal (bottom panel of Figure~\ref{fig:periodogram}), consistent with starspot-induced modulation. We therefore conclude that the star WASP-80 has a rotation period of $P_{\rm rot} = 3.15\pm0.01$ days.

\subsection{Gyrochronological Age and Independent Age Indicators}

\begin{figure}[h!]
    \centering
    \includegraphics[scale=0.3]{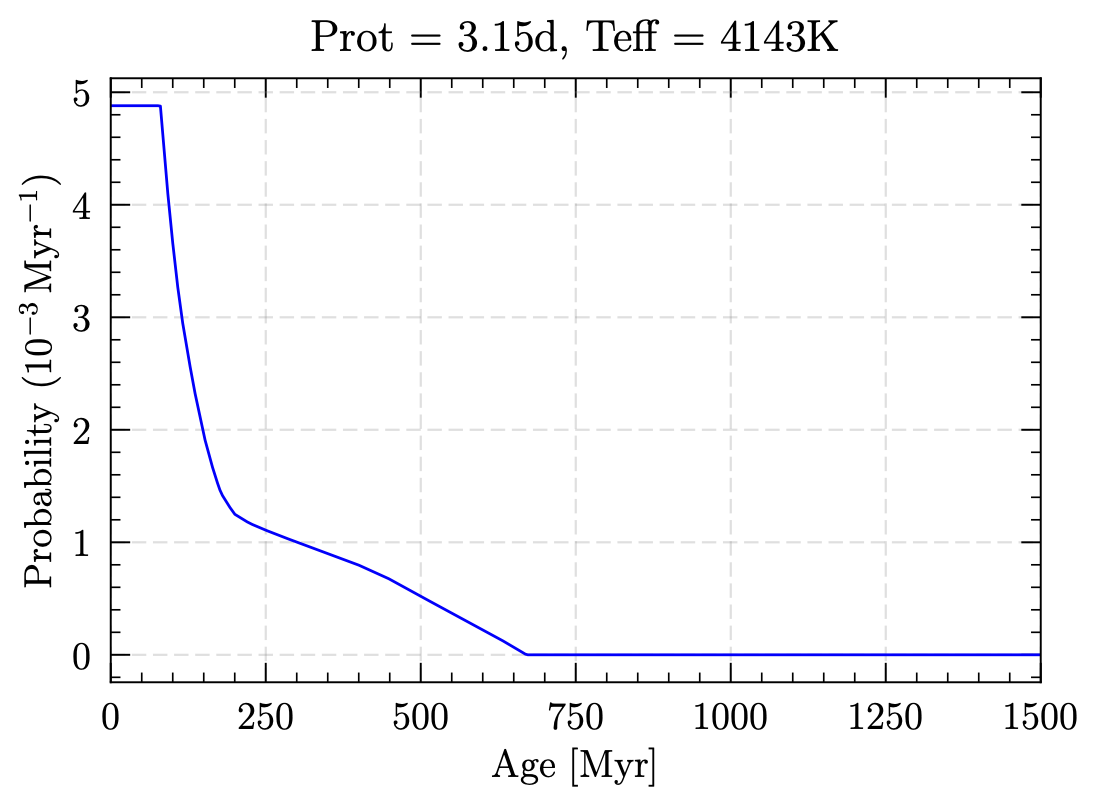}
    \caption{Posterior probability distribution for stellar age derived from gyrochronology for WASP-80 using \texttt{gyro-interp}, given a rotation period of $3.15\pm0.01$\,d and effective temperature of $4143\pm100$\,K.}
    \label{fig:gyro}
\end{figure}

Our primary age constraint comes from gyrochronology. Using the \texttt{gyro-interp} software tool \citep{Bouma2023} with our reported rotation period of WASP-80 ($P_{\mathrm{rot}} = 3.15 \pm 0.01$\,d), we estimate a 1$\sigma$ upper limit on the stellar age of $<195$\,Myr (68\% confidence). The posterior probability distribution corresponding to this estimate is shown in Figure~\ref{fig:gyro}.
Since the posterior remains high toward the youngest ages and does not exhibit a well-defined lower bound, we report a one-sided upper limit.
While several comoving companions have measured rotation periods from TESS photometry, most are M dwarfs that fall outside the effective temperature bounds where gyrochronological calibrations provide useful constraints. The two comoving K-type stars that do fall within the valid $T_{\mathrm{eff}}$ range yield consistent individual age estimates. 
A joint probability distribution analysis of these three stars results in an age constraint of \jointage{}\,Myr. However, we adopt the individual age upper limit of $<195$\,Myr described above for WASP-80 given the uncertainty in whether all kinematically identified companions are truly coeval and the improved rotation period precision available for WASP-80.

Independent age indicators for WASP-80 corroborate the star's young age. Using the \texttt{BAFFLES} software tool \citep{2020ApJ...898...27S} with archival HARPS photometry and the $\log(R'_{\mathrm{HK}})$ measurement \citep{Mancini2014}, we estimate a chromospheric age with a median of 153\,Myr (68\% CI: 44--850\,Myr).
Further, WASP-80's X-ray luminosity ($L_X = 8.0 \times 10^{28}$\,erg\,s$^{-1}$, \citep{King2018} ) places the target in the saturated regime, characteristic of early K-type stars younger than $\sim$300\,Myr. 
It is worth noting that while the original discovery paper \citep{Triaud2013} reports an equivalent width upper limit of $\mathrm{EW} < 30$\,m\AA{}, this measurement does not lend itself to providing a useful age constraint as late K-type dwarfs deplete photospheric lithium within $\sim$100\,Myr. No lithium equivalent width measurements were found for the comoving companions.

A previous age estimate for WASP-80 of $1352\pm222$\,Myr, derived using the \texttt{TATOO} tool \citep{gallet2020tatoo}, was adopted to calculate the bulk metallicity of WASP-80 b through a joint interior-atmospheric retrieval combining mass, age, and panchromatic JWST spectra \citep{acuna2025bulk}. That analysis yielded a high metallicity in their transmission-free retrievals, suggesting high internal temperatures inconsistent with the older age.
Our younger age estimate of $<195$\,Myr (68\% confidence) provides valuable context for understanding the orbital architecture and atmospheric composition of WASP-80 b.
The system's youth supports scenarios involving early-stage migration and dynamical evolution, informing interpretations of this gas giant's formation and migration history. 

\section{Atmospheric retrievals}\label{sec:retrievals}

For the 1D-RCPE atmosphere grid, we coupled the ScCHIMERA radiative-equilibrium solver \citep{piskorz2018ground, wiser2024lessons, wiser2025precise, iyer2023sphinx} with the Photochem chemical kinetics/photochemical solver \citep{wogan2025open}. The two codes were iterated until the maximum deviation in the temperature profile changed less than 10\%. We used the most recent reaction list on the VULCAN github page \citep{tsai2023photochemically} (converted to the Photochem format), which accounts for the updated sulfur chemistry pathways proposed by \citep{veillet2025inclusion}. We used the MUSCLES \cite{wilson2025mega} GJ 676 UV-spectrum as a proxy for WASP-80 and fixed the eddy diffusivity to $1\times 10^9 \mathrm{cm^2/s}$ and internal temperature to 150 K. We scaled the spectrum of GJ 676 so that it matched the measured XUV luminosity of WASP-80b, following the procedure described in \cite{fossati2022}. The grid was generated as a function of atmospheric metallicity ([M/H], uniformly scaling solar abundances \cite{lodders2009}) ranging from 0 ($1\times$) - 2.0 ($100\times$) dex in steps of 0.125, C/O (0.1 - 0.8 in steps of 0.1), and effective irradiation temperature (750 - 950 K, in steps of 25 K).
To fit the spectrum, the model atmospheres (temperature and gas mixing ratios as a function of pressure) were first packaged into an array as a function of [M/H], C/O, and $\mathrm{T_{irr}}$. Within the pymultinest sampler, we then interpolated the model atmospheres over the sampled grid-dimensions to then generate a transmission spectrum. Into this spectrum we also injected a gray cloud opacity, accounting for patchy clouds (two limbs), reference radius scaling, and offsets in the F322W2 and MIRI LRS datasets–for a total of 8 free parameters. The clouds were modeled as a vertically uniform gray cloud opacity (effectively an abundance weighted gray cross-section). We also experimented with power-law hazes and stellar heterogeneity, but these had virtually no impact on the nominal solutions, unsurprising as our data cover relatively long ($> 2.5 \mu$m) wavelengths where these effects are minimal. The opacity sources within the fitting procedure are previously described in \citep{Welbanks2024, wiser2025precise}, with the inclusion of CS$_2$ opacity generated from the latest HITRAN \citep{gordon2026hitran2024}.

\subsection{Testing The Significance of the CS\textsubscript{2} Feature near 4.6$\mu$m}

\begin{figure}
    \centering
    \includegraphics[width=0.8\linewidth]{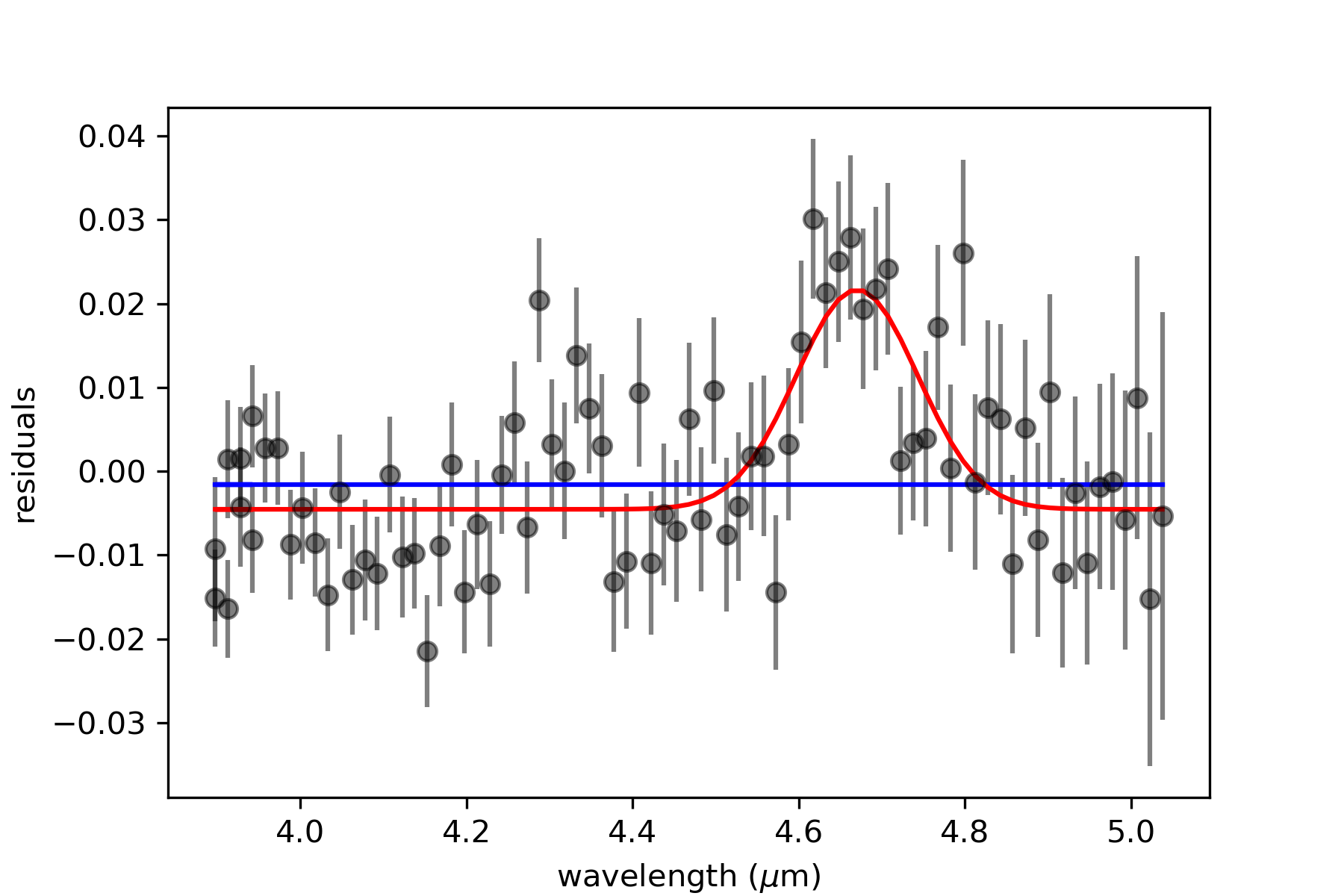}
    \caption{Residual data (black data points) after subtracting the best-fit forward model excluding the CS\textsubscript{2} contribution. In red is the best fit four-parameter Gaussian to the residual feature and in blue is the best fit single-parameter flat line to the residual feature.}
    \label{fig:quantify_signal}
\end{figure}

In addition to the detection significance provided by our retrieval analysis, we also wished to test the significance of the CS\textsubscript{2} feature near 4.6$\mu$m in a model-agnostic way. We did so following a method used previously to test the significance of the CO\textsubscript{2} in WASP-39b's atmosphere \citep{WASP39bERS_CO2}. Specifically, we recalculated the best fit transmission spectrum from the retrieval analysis with the CS\textsubscript{2} abundance set to zero, and then subtracted this from the observed data to create a residual spectrum in the neighborhood of the 4.6$\mu$m feature (Figure \ref{fig:quantify_signal}). We then fit this residual spectrum using a Gaussian model, and with a constant flat line model, and compared the Bayesian evidence to estimate which model was preferred by the data. We found that the Gaussian model was preferred at $\ln{B} = 28.2$, indicating that the 4.6$\mu$m feature is significantly present in the data itself.

\section{Tables}

\begin{table}[p]
\setlength{\arrayrulewidth}{1.3pt}
    \begin{tabular}{cc}
    	\hline
		\textbf{Orbital parameters for WASP-80b } & \textbf{Posteriors} \\
        \midrule
        P (days) & 3.067851917$\pm$0.000000007 \\
        T$_{\mathrm{C}}$ (BJD) & 2456487.4250142$\pm$0.000008\\
        $\log(\mathrm{a/R_{*}})$ & 1.10114$\pm$0.0001 \\
        $\cos(i)$ & 0.01867$\pm$0.00012 \\
        $\mathrm{R_p/R_*}$ (F322W2) & 0.17201$\pm$0.00003 \\
        $\mathrm{R_p/R_*}$ (F210M - simultaneous with F322W2) & 0.17201$\pm$0.00005 \\
        $\mathrm{R_p/R_*}$ (F444W) & 0.17147$\pm$0.00004 \\
        $\mathrm{R_p/R_*}$ (F210M - simultaneous with F444W) & 0.17203$\pm$0.00003 \\
        $\mathrm{R_p/R_*}$ (LRS) & 0.17093$\pm$0.00004\\
        \hline
\end{tabular}
\caption{The priors and the resulting posteriors of the orbital parameters from the joint broadband fit. The priors we use for the orbital parameters were taken from previous RV measurements \citep{triaud2015wasp}. Both F210M observations were taken simultaneously with the NIRCam spectroscopic observations and so we note which filter corresponds to which observation for their respective $\mathrm{R_p/R_*}$ value.}\label{tab:orbit}
\end{table}

\begingroup
\setlength{\tabcolsep}{12pt}   % Adjust horizontal space (default is 6pt)
\renewcommand{\arraystretch}{1.5}

\begin{table}[p]
\setlength{\arrayrulewidth}{1.3pt}
    \begin{tabular}{ccc}
    	\hline
		\textbf{Parameter} & \textbf{Estimate} & Prior (low, high)\\
        \midrule
        Redistribution & $1.21^{+0.06}_{-0.10}$ & (0.704, 1.304) \\
        $\mathrm{[M/H]}$ & $0.54^{+0.17}_{-0.12}$ & (0.0, 2.0)\\
        C/O & $0.43^{+0.12}_{-0.08}$ & (0.1, 0.8)\\
        $\log{(\kappa_\mathrm{cld})}$ & $-25.35^{+3.34}_{-2.58}$ & (-35, -20)\\
        $f_{\mathrm{cld}, t}$ & $0.36^{+0.06}_{-0.04}$ & (0, 1)\\
        $\times R_{p}$ & $1.06^{+0.01}_{-0.01}$ & (0.5, 1.5)\\
        $\mathrm{offset_{F322W2}}$ & $-27.87^{+16.53}_{-16.09}$ & (-300, 300) \\
        $\mathrm{offset_{LRS}}$ & $109.13^{+27.16}_{-31.19}$ & (-300, 300) \\
        $f_{\mathrm{spot}}$ & $0.01^{+0.01}_{-0.01}$ & (0, 0.5) \\
        $\mathrm{T_{spot}}$ & $3746^{+271}_{-620}$ & (2600, 4100)\\
        \hline
\end{tabular}
\caption{The parameters are the log metallicity relative to solar ($[\mathrm{M/H}]$), carbon-to-oxygen ratio (C/O), log cloud opacity in $\mathrm{cm^2/g}$ ($\log{(\kappa_\mathrm{cld})}$), the terminator cloud fraction ($f_{\mathrm{cld}, t}$), the offsets between modes (in ppm), and the starspot covering fraction ($f_{\mathrm{spot}}$, and the spot temperature ($\mathrm{T_{spot}}$).}
\label{tab:retrieval}
\end{table}

\endgroup

\clearpage

\section{Figures}\label{figures}

\section{Author contributions}

A.T. led the data analysis effort, the Pegasus analyses, verified the observing parameters, and led the writing of the paper. T.G.B. helped to plan the initial observations, contributed to the Pegasus analyses, led the photochemical modeling, and contributed to the text.  M.R.L. led the grid retrievals, contributed to the analysis and interpretation of the spectra, and contributed to the text. M.N. led the free retrievals, contributed to the analysis and interpretation of the spectra, and contributed to the text. T.J.B. contributed the Eureka! data analysis and the Eureka! stellar contamination analysis. E.S. contributed to observing specifications before the JWST launch and the tshirt data analysis. L.W. contributed to the text, conceptual direction of the paper, modeling analysis, and interpretation of the spectra. T.P.G. contributed to the scientific case for making the observations, led the observation planning and contributed to focusing the scientific content of the paper. M.S.F. led the gyrochronology and verified the stellar age estimate. J.J.F. helped to plan the initial observations and provided comments. M.R.L. contributed to the conceptual direction of the paper and interpretation of the spectra. K.O. helped to interpret the results and contributed to the text of the paper. N.M. ran the GCM models for the planet to provide physically consistent structure and spectra. S.M. used the PICASO atmospheric model to perform model fitting analysis on early versions of the spectra. M.M. provided insight in fitting the spectrum and gave comments on the paper. K.O. provided insight for interpreting the photochemical models and gave comments on the paper. V.P. helped with the physical interpretation of the spectrum and interpreting the photochemical models. Y.R. contributed text to the draft, provided interpretations of the modeled spectrum, and contributed to the conceptual direction of the paper. L.S.W. provided the results of the emission spectrum and conceptual direction of the paper.

\section{Acknowledgments}

We thank Marcia Rieke for providing time from the NIRCam GTO allocation to make the NIRCam observations presented and analyzed here (from JWST GTO program 1185). T.P.G. acknowledges funding support from the Next Generation Space Telescope Flight Investigations program (now JWST) via NASA WBSs 411672.07.04.01.02 and 411672.07.05.05.03.02. This research has made use of the Astrophysics Data System, funded by NASA under Cooperative Agreement 80NSSC21M00561. This research has made use of the NASA Exoplanet Archive, which is operated by the California Institute of Technology, under contract with the National Aeronautics and Space Administration under the Exoplanet Exploration Program. The data were obtained from the Mikulski Archive for Space Telescopes at the Space Telescope Science Institute, which is operated by the Association of Universities for Research in Astronomy, Inc., under NASA contract NAS 5-03127 for JWST.
\clearpage

\bibliography{sn-bibliography}% common bib file

\end{document}